\documentclass[a4paper]{article}

\usepackage{spconf}
\usepackage{epsfig,amssymb,amsmath,multirow,boldline,graphicx,makecell,hyperref}
\usepackage{kotex}

\def\bmu{{\bm \mu}}

\def\z{{\bm z}}
\def\Z{{\bm Z}}
\def\x{{\bm x}}
\def\X{{\bm X}}

\usepackage{xcolor}

\title{Unsupervised Representation Learning of Speech \\for Dialect Identification}
\name{Suwon Shon, Wei-Ning Hsu, James Glass }
\address{Computer Science and Artificial Intelligence Laboratory \\
Massachusetts Institute of Technology \\
Cambridge, MA 02139 USA \\
{\small\texttt{\{swshon,wnhsu,glass\}@mit.edu}}}

\begin{document}

\maketitle
\begin{abstract}
In this paper, we explore the use of a factorized hierarchical variational autoencoder (FHVAE) model to learn an unsupervised latent representation for dialect identification (DID). An FHVAE can learn a latent space that separates the more static attributes within an utterance from the more dynamic attributes by encoding them into two different sets of latent variables. Useful factors for dialect identification, such as phonetic or linguistic content, are encoded by a segmental latent variable, while irrelevant factors that are relatively constant within a sequence, such as a channel or a speaker information, are encoded by a sequential latent variable. The disentanglement property makes the segmental latent variable less susceptible to channel and speaker variation, and thus reduces degradation from channel domain mismatch. We demonstrate that on fully-supervised DID tasks, an end-to-end model trained on the features extracted from the FHVAE model achieves the best performance, compared to the same model trained on conventional acoustic features and an i-vector based system. Moreover, we also show that the proposed approach can leverage a large amount of unlabeled data for FHVAE training to learn domain-invariant features for DID, and significantly improve the performance in a low-resource condition, where the labels for the in-domain data are not available.
% In this paper, we explore the use of a factorized hierarchical variational autoencoder (FHVAE) model to learn an unsupervised latent representation for dialect identification (DID). An FHVAE can learn a latent space that separates the more static attributes within an utterance (e.g., speaker and channel) from the more dynamic attributes (e.g., phonetic) by encoding them into two different sets of latent variables. Useful factors for dialect identification, such as phonetic or linguistic content, are encoded by a segmental latent variable, while irrelevant factors that are relatively constant within a sequence, such as a channel or a speaker information, are encoded by a sequential latent variable. The disentanglement property makes the segmental latent variable less susceptible to channel and speaker variation, and thus reduces degradation from channel domain mismatch. We demonstrate that on fully-supervised DID tasks, an end-to-end model trained on the features extracted from the FHVAE model achieves the best performance, compared to the same model trained on conventional acoustic features and an i-vector based system. Moreover, we also show that the proposed approach can leverage a large amount of unlabeled data for FHVAE training to learn domain-invariant features for DID, and significantly improve the performance in a low-resource condition, where the labels for the in-domain data are not available.
\end{abstract}

\section{Introduction}

Over the last few years, combinations of i-vectors and Deep Neural Networks (DNNs)~\cite{Cardinal2015, Richardson2015, Dehak2011b} have achieved state-of-the-art results for speaker recognition and language identification (LID). DNN-based end-to-end systems~\cite{Jin2017, Trong2016, Nagraniy2017, DavidSnyder2016} have recently obtained comparable or slightly better performance on these tasks. In comparison to LID, Dialect Identification (DID) is a relatively unexplored task because DID is often regarded as a pecial case of LID.
% One of the main reasons is because of a lack of common datasets, while another reason is that DID is often regarded as a special case of LID, so researchers tend to concentrate on the more general problems of language and speaker recognition. 
However, DID is, in fact, a much more challenging task compared to LID, due to the high similarity between dialects~\cite{Ali2016,Khurana2017,Maryam2018,Zampieri2018}.

In \cite{Shon2018a}, the authors investigated an end-to-end approach to a DID task using an Arabic dialect dataset. The authors applied a Convolutional Neural Network (CNN) and conducted extensive experiments on comparing the use of different acoustic features as well as data augmentation methods, which demonstrates the importance of feature selection and the effectiveness of increasing the amount of data. However, such methods suffer from severe performance degradation under domain mismatch condition, where training and testing data are drawn from different domains. Unfortunately, this is not a rare condition due to the scarcity of such datasets. Thus, learning a better speech representation that is domain invariant becomes essential for DID. In previous studies, domain mismatch was mainly addressed in the context of speaker recognition\cite{Aronowitz2014,Garcia-Romero2014b,Shum2014,shon2017,Shon2017a}. Among them, Inter Dataset Variability Compensation (IDVC) is a simple but powerful approach when in-domain data are presented but unlabeled. We will also investigate this approach on dialect identification. Another approach that can be applied to this condition was studied by Zhang and Hansen~\cite{zhang2018language, Zhang2017inter}. They presented general Autoencoder approach and unsupervised bottleneck feature (uBNF) extraction approach to language/dialect identification task. Because both approaches do not need ground truth label, they use a big amount of unlabeled data to compensate domain mismatches. They concluded uBNF achieved the best performance on the tasks. We conducted uBNF extraction on our experimental condition for performance comparison.

In this paper, we present unsupervised representation learning of dialectal speech using a Factorized Hierarchical Variational Autoencoder (FHVAE)~\cite{hsu2017unsupervised}. An FHVAE is a variant of variational autoencoders (VAEs)~\cite{kingma2013auto}, which models a generative process of sequential data with a hierarchical graphical model, and defines a corresponding inference model for variational inference. Similar to VAEs, an FHVAE is trained to maximize a lower bound of the marginal likelihood, and hence does not require any supervision, which enables utilization of an unlabeled in-domain dataset for representation learning. In particular, such a model can represent static and dynamic generating factors within an utterance with a latent sequence variable and latent segment variable respectively. As shown in~\cite{hsu2017unsupervised}, when an FHVAE model is trained on speech data, channel and speaker-related information will be encoded in the latent sequence variable, while language related information like accent, tone, rhythm are encoded in the latent segment variable. Therefore, if we use the latent segment variable as a new feature for dialect identification, the system could be more robust to a domain and speaker variability. We incorporated this latent segment variable into our end-to-end dialect identification system. 

We conducted experiments on two different conditions where labels of in-domain data are only available in one setting and the other is not. I-vector and end-to-end systems trained on traditional acoustic features also served as a baseline result. The results indicate that using the latent segment representation from the FHVAE model to train an end-to-end DID system outperforms all other baselines and suffers less degradation when a labeled in-domain dataset is not available.

\section{Dialect/Language Identification System}
\subsection{I-vectors and Bottleneck Features}
The i-vector has been regarded as the state-of-the-art representation for speaker and language identification, especially when combined with a DNN for i-vector extraction. The most successful approach uses a DNN to generate bottleneck (BN) features from an acoustic model~\cite{Cardinal2015, Matejka2014, Richardson2015, Khurana2017, Ali2016}. The acoustic model can be trained using DNNs with state-level alignment information, where the DNN has one relatively constricted layer with a small number of hidden units (i.e., a "bottleneck").
The bottleneck layer activation can be used as a new acoustic feature for training the i-vector extractor, or it can be fed into a second stage DNN acoustic model to produce a Stacked Bottleneck (SBN) feature~\cite{Matejka2014, Zhang2014}. Although the acoustic model is often trained on monolingual data (e.g., English), bottleneck features perform reasonably well on LID tasks~\cite{Matejka2014}. For DID, since the dialects are from the same language family, using an SBN that is extracted from a subset of the dialects can achieve excellent performance~\cite{Shon2017c}. Training an i-vector extractor is exactly the same for a speaker recognition or an LID system. The only difference is learning the subsequent projections.
% \wnhcomment{Q: what projection?} \swscomment{because the i-vector is trained without supervision, we need to project the i-vectors to task oriented i-vector like speaker or language(usually using LDA)} that are tailored to the specific task.

\subsection{End-to-End CNN/DNN System}
Recently, end-to-end approaches have achieved impressive performance compare to conventional i-vector approach for both LID~\cite{Jin2017,Trong2016,Shon2018a,Snyder2018} and speaker recognition ~\cite{DavidSnyder2016,Nagraniy2017,Shon2018frame}. In \cite{Shon2018a}, the authors conducted detailed experiments on an end-to-end system using a dataset augmentation approach with acoustic features ranging from Mel-Frequency Cepstral Coefficients (MFCCs) to spectrograms. The end-to-end system performed significantly better than i-vectors if the training dataset had a large amount of diversity. 

In this work, we adopt the end-to-end system proposed in~\cite{Shon2018a}.
This system has a stack of CNN layers, followed by a global pooling layer that aggregates frame level representations to produce to utterance level representations. The output of global pooling layer is followed by two fully connected layers. Specifically, the network consists of four 1-dimensional CNN layers (40$\times$5 - 500$\times$7 - 500$\times$1 - 500$\times$1 filter sizes; with 1-2-1-1 strides; the number of filters is 500-500-500-3000) and two fully connected layers (1500-600). The size of the final softmax layer is determined by the task-specific speaker or language labels and the softmax output can be used directly as a score for each dialect class for the DID task. We considered MFCCs and Mel-Filterbank energies (FBANK) as inputs to the end-to-end system.

\section{Unsupervised Learning using FHVAE}
% \textbf{need further revision and extension/add graphical model figure/connect z1 to dialect}
% \textit{Description of FHVAE (detailed setup information for training should be placed at section 5.1)}

In this section, we introduce an FHVAE model for unsupervised representation learning from dialectal speech and explain how we extract features from such models for dialect identification.

\begin{figure}[t]
\centerline{\includegraphics[width=.8\linewidth]{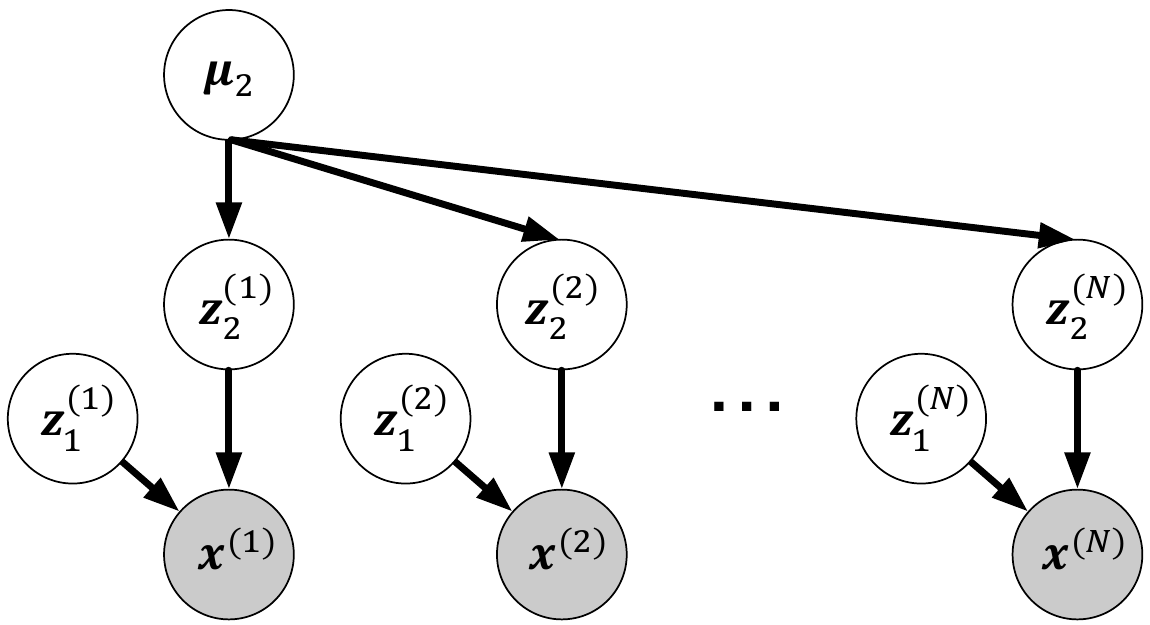}}
\caption{Graphical illustration of the FHVAE generative model. Grey nodes denote the observed variables, and white nodes are the latent variables.}
\label{fig:fhvae_gen}
\end{figure}

An FHVAE~\cite{hsu2017unsupervised} is a variant of variational autoencoders~\cite{kingma2013auto,hsu2017learning} that models a probabilistic hierarchical generative process of sequential data, and learns disentangled and interpretable representations.
Generation of a sequence of $N$ segments, $\X = \{ \x^{(n)} \}_{n=1}^N$ involves one sequence-level latent variable: $\bmu_2$, and $N$ pairs of segment-level latent variable: $\z_1$ and $\z_2$, as follows:
  \begin{enumerate}
    \item a \textit{s-vector} $\bmu_2$ is drawn from $p(\bmu_2) = \mathcal{N}(\bmu_2 | \bm{0}, \sigma^2_{\bmu_2}\bm{I})$.
    \item $N$ i.i.d. \textit{latent segment variables} $\bm{Z}_1 = \{ \bm{z}_1^{(n)} \}_{n=1}^N$ are drawn from a global prior $p(\bm{z}_1) = \mathcal{N}(\bm{z}_1 | \bm{0}, \sigma^2_{\bm{z}_1}\bm{I})$.
    \item $N$ i.i.d. \textit{latent sequence variables} $\bm{Z}_2 = \{ \bm{z}_2^{(n)} \}_{n=1}^N$  are drawn from a sequence-dependent prior $p(\bm{z}_2 | \bmu_2) = \mathcal{N}(\bm{z}_2 | \bmu_2, \sigma^2_{\bm{z}_2}\bm{I})$.
    \item $N$ i.i.d. sub-sequences $\bm{X} = \{ \bm{x}^{(n)} \}_{n=1}^N$ are drawn from $p(\bm{x} | \bm{z}_1, \bm{z}_2) = \mathcal{N}(\bm{x} | f_{\mu_x}(\bm{z}_1, \bm{z}_2), diag(f_{\sigma^2_x}(\bm{z}_1, \bm{z}_2)))$, where $f_{\mu_x}(\cdot,\cdot)$ and $f_{\sigma^2_x}(\cdot,\cdot)$ are parameterized by a decoder neural network.
  \end{enumerate}
We illustrate this process in Figure~\ref{fig:fhvae_gen}.
By imposing a sequence-dependent prior to $\z_2$, the model is encouraged to represent with $\z_2$ the generating factors that are relatively consistent within a sequence.
For example, such factors can include microphone frequency response, room impulse response, and general vocal tract characteristics particular to a speaker.  On the other hand, $\z_1$ tends to encode information about the residual generating factors that change from segment to segment, such as phonetic/linguistic content.

Since the exact posterior is intractable, an FHAVE approximates the true posterior with an amortized inference model $q(\bmu_2 | \X) \prod_{n=1}^N q(\bm{z}_1^{(n)} | \bm{x}^{(n)}, \bm{z}_2^{(n)}) q(\bm{z}_2^{(n)} | \bm{x}^{(n)})$. 
Specifically, both $q(\bm{z}_1 | \bm{x}, \bm{z}_2)$ and $q(\bm{z}_2 | \bm{x})$ are Gaussian distributions, whose mean and variance are parameterized by two encoder neural networks with the variables they are conditioned on as inputs.
Specifically, both $q(\bm{z}_1 | \bm{x}, \bm{z}_2)$ and $q(\bm{z}_2 | \bm{x})$ are Gaussian distributions. 
Each of them is parameterized by one encoder neural network that predicts the posterior mean and variance, with the variables it is conditioned on as inputs.
On the other hand, $q(\bmu_2 | \X)$ is a fixed-variance Gaussian distribution, whose mean is parameterized with a maximum a posterior (MAP) estimation $\sum_{n=1}^N \hat{\bm{z}}_2^{(n)} / (N + \sigma^2_{\bm{z}_2} / \sigma^2_{\bmu_2})$ for testing, where $\hat{\bm{z}}_2^{(n)}$ denotes the posterior mean of $\z_2$ for $\x^{(n)}$.
For training, a trainable lookup table of $\bmu_2$ posterior mean for each training sequence is used instead.
which enables optimization at the segment level, and utilization of the discriminative loss proposed in~\cite{hsu2017unsupervised} to encourage disentanglement.

In this work, we pool the in-domain and out-of-domain dialectal speech data to train an FHVAE, in order to learn a disentangled and interpretable representation.
Note that since training of FHVAE models is unsupervised, we can actually apply the FHVAE model even though the dataset has no dialect label.
In addition, because at the acoustic level, variation in dialects correlates with phonetic and lexical variability, but not channel response or vocal tract characteristics, we argue that useful information regarding dialect identification is actually encoded in $\z_1$ instead of $\z_2$. 
Hence, similar to~\cite{hsu2018extracting}, we extract $\Z_1$ for each sequence as the new feature representation and use it for training DNN-based end-to-end systems.
After an FHVAE model is trained, we extract $\Z_1$ for each sequence and use it 
Among the disentangled generating factors, we hypothesize that useful information regarding dialect identification is encoded in $\z_1$

\section{Domain Mismatched Dialectal Speech}
The MGB-3 dataset partitions are shown in Table~\ref{tab:data}. Each partition consists of five Arabic dialects: EGY, LEV, GLF, NOR, and MSA.  Detailed corpus details can be found in~\cite{Ali2017}. Although the development set is relatively small compared to the training set, it matches the test set channel conditions, and thus provides valuable information about the test domain. In the following experiment, we divided the train and development set into a small subset and limit the use of labels on the development set to simulate a low-resource condition.

\begin{table}[b]
\centering
\caption{MGB-3 Dialectal Arabic Speech Dataset Properties.}
\label{tab:data}
\resizebox{0.9\linewidth}{!}{%
\begin{tabular}{c||c|c|c}
\hlineB{2}
\begin{tabular}[c]{@{}c@{}}Dataset\end{tabular}& \begin{tabular}[c]{@{}c@{}}Training\end{tabular}& \begin{tabular}[c]{@{}c@{}}Development\end{tabular}   & \begin{tabular}[c]{@{}c@{}}Test\end{tabular}  \\ \hlineB{2}
Utterances & 13,825 & 1,524 & 1,492\\  \hline
Size & 53.6 hrs & 10 hrs & 10.1 hrs\\  \hline
\begin{tabular}[c]{@{}c@{}}Channel\\ (recording)\end{tabular}& \begin{tabular}[c]{@{}c@{}}Carried out\\ at 16kHz\end{tabular} & \multicolumn{2}{c}{\begin{tabular}[c]{@{}c@{}}Downloaded directly from\\ a high-quality video server\end{tabular}} \\ \hlineB{2}
\end{tabular}%
}
\end{table}

\begin{table}[t]
\centering
\caption{Baseline accuracy performance when using train and development dataset for training.}
\label{tab:perform_baseline}
\resizebox{1.0\linewidth}{!}{%
\begin{tabular}{c||c|c}
\hlineB{2}
\multirow{2}{*}{System} & \multicolumn{2}{c}{\begin{tabular}[c]{@{}c@{}}Accuracy on Test set\end{tabular}} \\ \cline{2-3} 
 & If Dev. set is labeled & If Dev. set is unlabeled\\ \hlineB{2}
I-vector  & 57.44 & 46.11\\ \hline
End-to-End (MFCC)  & \textbf{65.55} & \textbf{48.86}\\ \hline
End-to-End (FBANK) & 64.81 & 47.11 \\ \hlineB{2}
%BNF i-vector* & 60.32 & \textbf{55.29}\\ \hlineB{2}
\end{tabular}%
}
\end{table}

\section{Experiment}
\subsection{Experiment setup}
For i-vector extraction, MFCCs were used to generate 60-dimensional acoustic features which consist of 20 MFCCs and their delta and delta-delta's.  Cepstral Mean Normalization was used for feature normalization. A GMM-UBM was trained using MFCCs with 2,048 mixture components, then a Total Variability (TV) matrix was trained to extract 600-dimensional i-vectors. Since the GMM-UBM and TV can be trained without labels, we used training and development dataset for all experimental conditions in next sections. A Support Vector Machine (SVM) was used to measure the similarity between a test utterance and 5 dialects~\cite{Ali2016}. We did not consider a bottleneck-feature (BNF) based i-vector system for comparison. A BNF extractor must be trained with supervision using additional data which have phoneme alignments. Using this extra information was not comparable to any system using only the MGB-3 data.

For the end-to-end DID system, we used MFCC and FBANK features. To extract the features, a spectrogram was computed using a 400 sample FFT window length with 160 sample advance which is equivalent to 25ms window and 10ms frame-rate for 16kHz audio. A total of 40 coefficient were extracted for both features and then normalized to have zero mean and unit variance. The DNN structure is the same as~\cite{Shon2018a}, with four CNN and 2 FC layers as described in Section 2. The stochastic gradient descent (SGD) learning rate was 0.001 with a decay every 50,000 mini-batches with a factor of 0.98. Rectified Linear Units (ReLUs) were used for activation nonlinearities. We used a different dataset for training the network considering various experimental condition since the end-to-end DNNs need dialect label.

To train FHVAE models, we let each segment $\x$ be 20 frames of FBANK features.
Following the setting in~\cite{hsu2017unsupervised}, we set $\sigma^2_{\z_1}=\sigma^2_{\bmu_2}=1$, $\sigma^2_{\z_2}=0.25$, and dimensions of $\z_1$ and $\z_2$ to be both 32.
We configured the two encoders and the decoder to be two-layer LSTM\cite{hochreiter1997long} networks with 256 memory cells, followed by affine transform layers predicting mean and log variance of corresponding variables, similar to the architecture used in~\cite{hsu2018extracting}.
FHVAE models are trained to maximize the discriminative segment variational lower bound proposed in~\cite{hsu2017unsupervised} with a discriminative weight $\alpha = 10$.
%   In addition, regular FHVAE training is not scalable to hundreds thousands of utterances; instead, we use hierarchical sampling-based training algorithm proposed in~\cite{} with batches of 5,000 utterances.
  Adam~\cite{kingma2014adam} with $\beta_1 = 0.95$ and $\beta_2 = 0.999$ is used to optimize all models.
  Tensorflow~\cite{abadi2016tensorflow} is used for implementation.

The performance was measured in accuracy, Equal Error Rate (EER) and minimum decision cost function C\textsubscript{avg}*100. Accuracy was measured by taking the dialect showing the maximum score between a test utterance and the 5 dialects. Minimum C\textsubscript{avg}
*100 was computed from hard decision errors and a fixed set of costs and priors from~\cite{lre15}.

\subsection{Resource Limitation Impact on Domain Mismatch}
We compared the performance of the baseline approaches in Table~\ref{tab:perform_baseline}. When using the training and development set including their label, the end-to-end system shows impressive performance, since the approach is powerful when the training and test conditions are the same.
However, when the development set has no labels, all three baselines show similar accuracy and the end-to-end system no longer has an advantage compared to the traditional i-vector approach. 
It is interesting that i-vector does not have advantage on this case although UBM is trained using both train and development set since the model does not need supervision.
This analysis implies that although the development dataset is only 10 hours long, averaging 2 hours for each dialect, it carries valuable information about the target domain and has a powerful impact on the performance. Also, it is observed that GMM-UBM is not efficient to learn domain mismatched information when the in-domain set is very small. Both i-vector and end-to-end systems are unable to use this valuable information properly due to lack of labels and accuracy subsequently degraded about 20\% and 25\% respectively compared to when there are labels. We conducted more detailed performance comparison experiments on the two condition in the next section.

\begin{table}[t]
\centering
\caption{Performance on resource-rich condition (development set with labels).}
\label{tab:perform_trndev}
\resizebox{1.0\linewidth}{!}{%
\begin{tabular}{c||c|c|c}
\hlineB{2}
 & Accuracy & EER & C\textsubscript{avg}*100 \\ \hlineB{2}
i-vector & 57.44 & 24.43 & 23.79 \\ \hline
End-to-end (MFCC) & 65.55 & 20.24 & 19.92 \\ \hline
End-to-end (FBANK) & 64.81 & 20.22 & 19.91 \\ \hline
\textbf{End-to-end (FHVAE\_$\z_1$)} & \textbf{67.98} & \textbf{18.62} & \textbf{18.32} \\ \hline
End-to-end (FHVAE\_$\z_2$) & 54.55 & 27.39 & 27.35 \\ \hlineB{2}
\end{tabular}%
}

\end{table}

\begin{table}[t]
\centering
\caption{Performance on resource-poor condition (development set without labels).}
\label{tab:perform_trn}
\resizebox{1.0\linewidth}{!}{%
\begin{tabular}{c||c|c|c}
\hlineB{2}
 & Accuracy & EER & C\textsubscript{avg}*100 \\ \hlineB{2}
i-vector & 46.11 & 32.77 & 32.08 \\ \hline
End-to-end (MFCC) & 48.86 & 29.31 & 28.61 \\ \hline
End-to-end (FBANK) & 47.86 & 30.19 & 29.67 \\ \hline
\textbf{End-to-end (FHVAE\_$\z_1$)} & \textbf{58.16} & \textbf{25.40} & \textbf{24.66} \\ \hline
End-to-end (FHVAE\_$\z_2$) & 36.36 & 39.00 & 38.32 \\ \hlineB{2}
\end{tabular}%
}

\end{table}

\subsection{Resource-Rich Condition}
Consider the resource-rich condition in which there are labels on the in-domain data. In this condition, we can fully utilize the development set with dialect labels as part of training. 
All end-to-end approaches except FHVAE\_$\z_2$ show significantly better performance than i-vector approaches on all measurements. Using $\z_1$ of FHVAE is the best system and $\z_2$ of FHVAE is the worst as shown in Table~\ref{tab:perform_trndev}. 
As we expected, linguistic information to distinguish language is encoded at $\z_1$, the latent segment variable, and shows better performance than other features. 
On the other hand, because the latent sequence variable $\z_2$ carries information that is not directly related to the dialect identity, such as channel, it is not surprising that the model trained on $\z_2$ has the worst performance.
% Since the $\z_2$, latent sequence variable, shows worst performance, it seems that segment-invariant \wnhcomment{(comment: segment-invariant is a bit confusing)} attributes such as speaker or channel domain information was well separated from the original speech by FHVAE and was encoded by the $\z_2$ variable.

\begin{figure*}[t]
\centerline{\includegraphics[width=21.5cm]{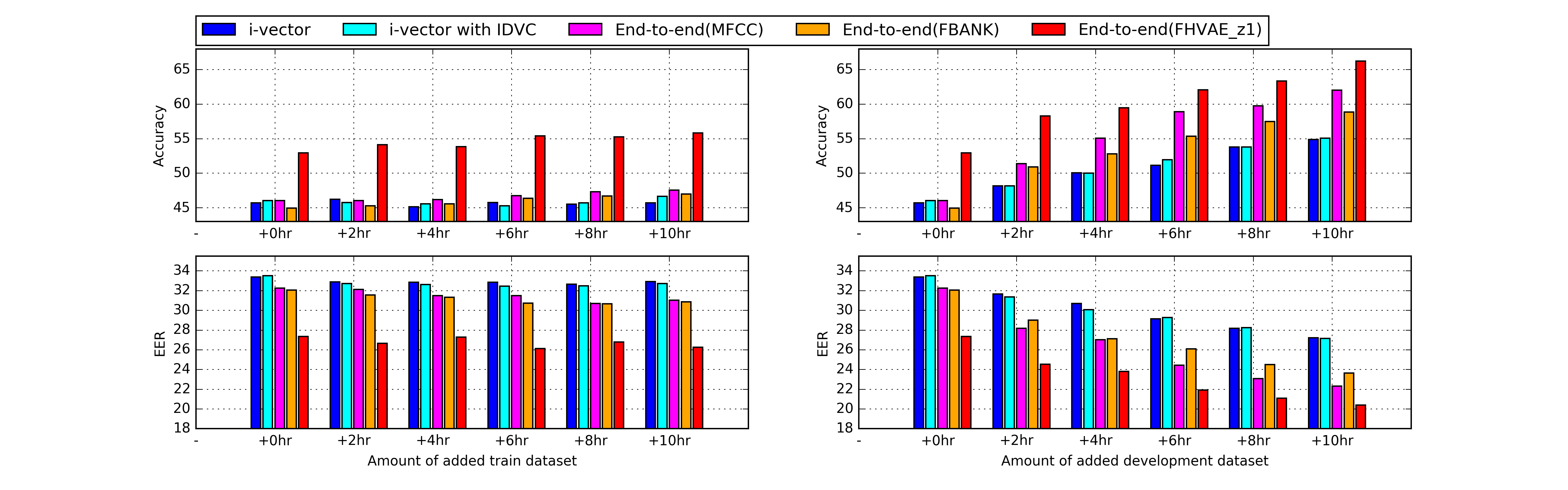}}
\caption{Performance variation with incremental train (mismatched domain) and development data (matched domain)}
\label{fig:perform_addtrn}
\end{figure*}

\subsection{Resource-Poor Condition}
As a fully-labeled in-domain dataset is not always available or can be difficult to collect, we consider a resource-poor condition, where an in-domain dataset is available but without any labels. For this condition, an i-vector extractor and GMM-UBM are trained without supervision, so the in-domain data could be used for training even though it is a small amount. However, the end-to-end system is unable to use the in-domain data because training must be done in a supervised way. As for the resource-poor condition, using $\z_1$ of FHVAE shows the best performance on all measurements as shown in Table~\ref{tab:perform_trn}. It is interesting that the performance is very similar to that achieved by the i-vector system on the resource-rich condition.

Apart from the absolute numbers in the tables, the greater performance degradation due to lack of in-domain labels implies that the feature or system is more dependent on the domain. The FHVAE\_$\z_1$ based end-to-end approach accuracy degraded only 14\% compared to the resource-rich condition, but the other methods degraded about 20\%-26\%. Particularly, FHVAE\_$\z_2$ degraded by 33\% which implies that the $\z_2$ variable is significantly domain-related as we argued previously.

\begin{table}[t]
\centering
\caption{Division of Training and Development Dataset.}
\label{tab:divided_data}
\resizebox{\linewidth}{!}{%
\begin{tabular}{c||c|c|c|c|c|c}
\hlineB{2}
Dataset    & Trn-30h & Trn-1 & Trn-2 & Trn3  & Trn-4 & \multicolumn{1}{c}{Trn5} \\ \hlineB{2}
Utterances & 7441      & 554     & 511     & 518     & 548    & \multicolumn{1}{c}{549}    \\ \hline
Size       & 30hrs  & 2hrs  & 2hrs  & 2hrs  & 2hrs  & \multicolumn{1}{c}{2hrs} \\ \hline\hline
Dataset    & Dev-1  & Dev-2 & Dev-3 & Dev-4 & Dev-5 &                           \\ \cline{1-6} 
Utterances & 308      & 303     & 322     & 318     & 315     &                           \\ \cline{1-6}
Size       & 2hrs   & 2hrs  & 2hrs  & 2hrs  & 2hrs  &                           \\ \cline{1-6} \hlineB{2}
\end{tabular}%
}

\end{table}

\begin{table}[t]
\centering
\caption{Dataset usage by system. $\alpha$ is additional labeled set that can be added from \{Trn-1$\sim$5, Dev-1$\sim$5\} from Table~\ref{tab:divided_data}.}
\label{tab:dataset_usage}
\resizebox{\linewidth}{!}{%
\begin{tabular}{c||c|c}
\hlineB{2}
System & \begin{tabular}[c]{@{}c@{}}Unlabeled data\\ (Train + Dev.)\end{tabular} & \begin{tabular}[c]{@{}c@{}}Labeled data\\ (Train-30h + $\alpha$)\end{tabular} \\ \hlineB{2}
i-vector & UBM, TV & SVM \\ \hline
End-to-end (MFCC, FBANK) & - & End-to-end model \\ \hline
End-to-end (FHVAE) & FHVAE model & End-to-end model \\ \hlineB{2}
\end{tabular}%
}
\end{table}

To examine the efficiency of the proposed approach, we partition the train and development datasets into smaller subsets to have specific amounts of data, as shown in Table~\ref{tab:divided_data}. While all speech from the train and development sets are available, we only used labels for 30 hours of speech from the train set as an ``essential" system, and gradually added from 2 to 10 hours of additional labels from the train and development sets in 2 hour increments. Since each system has a different algorithm, we specified dataset usage in Table~\ref{tab:dataset_usage}. For example, to add 6 hours of labels from the development set, we used Dev-1, Dev-2 and Dev-3 labels from Table~\ref{tab:divided_data}. The result is shown in Figure~\ref{fig:perform_addtrn}. We also applied IDVC, which can be applied to the i-vector system for domain compensation. 

When adding the train set, the systems show slight or no improvement except the end-to-end system using the FHVAE $\z_1$ variable. For both conditions, the addition of IDVC does not substantially improve the baseline i-vector results.  When incrementally providing an additional 2 hours of the train dataset, the improvement on accuracy and EER is under 1\% for all systems. However, when incrementally adding 2 hours of the dev dataset, the accrual is between 4\% and 7\% for the i-vector based and end-to-end approaches respectively. In particular, the gap between MFCC based and $\z_1$ of the FHVAE based end-to-end approach is getting closer when adding in-domain (development) set while they keep the same separation when adding out-of-domain (train) set. This observation tells us that if we did not decide on a target domain or did not know about the target domain, we can extract domain invariant features using FHVAE by adding large quantities of unlabeled data from various domains. And it also indicates that if we know the target domain precisely, and it is possible to obtain a target domain dataset with rich labels,  using MFCCs or raw feature on the end-to-end system can be optimized to the specific domain and gives a comparable performance to the $\z_1$ FHVAE system.
\subsection{Performance comparison with uBNF}
In this section, we compared uBNF feature~\cite{zhang2018language, Zhang2017inter} which was extracted from DNN trained in an unsupervised manner. The approach was successfully adopted on DID task by modifying network structure considering the relatively small size of dataset~\cite{Bulut2017}. We trained GMM-UBM model with same parameter as~\cite{Bulut2017}. Using the posterior label estimation, 4-layer DNN was trained to extract 40-dimensional uBNF feature. The extracted uBNF feature was used to train the end-to-end DID system in both resource-rich and poor condition. In both conditions, the proposed FHVAE\_$z_1$ shows better performance in all indexes. A possible reason for this difference is that the uBNF obtain the ground truth label from GMM-UBM, the performance is dependent on the GMM-UBM. But the FHVAE does not depend on such as label obtained from another unsupervised learning, it has more advantage to learning the speech unsupervised manner. 

\begin{table}[t]
\centering
\caption{Performance comparison on MGB-3 test set.}
\label{tab:perform_compare}
\resizebox{1.0\linewidth}{!}{%
\begin{tabular}{c||c|c|c}
\hlineB{2}
Resource-poor & Accuracy & EER & C\textsubscript{avg}*100 \\ \hlineB{2}
End-to-end (uBNF~\cite{zhang2018language}& 56.64 & 27.46 & 26.92 \\ \hline
\textbf{End-to-end (FHVAE\_$\z_1$)} & \textbf{58.16} & \textbf{25.40} & \textbf{24.66} \\ \hline\hline
Resource-rich & Accuracy & EER & C\textsubscript{avg}*100 \\ \hline
End-to-end (uBNF~\cite{zhang2018language}) & 66.24 & 19.98 & 19.63 \\ \hline
\textbf{End-to-end (FHVAE\_$\z_1$)} & \textbf{67.98} & \textbf{18.62} & \textbf{18.32} \\ \hlineB{2}
\end{tabular}%
}
\end{table}

\section{Conclusion}
In this paper, we describe domain invariant features from unsupervised learning of dialectal speech. The feature, a latent segmental variable, can be encoded by an FHVAE. We investigated the proposed approach along with several baselines such as i-vectors and end-to-end methods based on CNNs using conventional acoustic features. The experiments explored two scenarios, whether the in-domain dataset has a dialect label or not, in order to explore the effectiveness of unsupervised learning of dialectal speech in various domains. From the experiments, we observed that the proposed approach is able to separate segmental and sequential level information that generalize better to new domains.  While the proposed approach shows significant improvement in all conditions, we verified that it has a greater advantage in the case where a large amount of unannotated audio is available. 

\bibliographystyle{IEEEbib}

\bibliography{mybib}

% \begin{thebibliography}{9}
% \bibitem[1]{Davis80-COP}
%   S.\ B.\ Davis and P.\ Mermelstein,
%   ``Comparison of parametric representation for monosyllabic word recognition in continuously spoken sentences,''
%   \textit{IEEE Transactions on Acoustics, Speech and Signal Processing}, vol.~28, no.~4, pp.~357--366, 1980.
% \bibitem[2]{Rabiner89-ATO}
%   L.\ R.\ Rabiner,
%   ``A tutorial on hidden Markov models and selected applications in speech recognition,''
%   \textit{Proceedings of the IEEE}, vol.~77, no.~2, pp.~257-286, 1989.
% \bibitem[3]{Hastie09-TEO}
%   T.\ Hastie, R.\ Tibshirani, and J.\ Friedman,
%   \textit{The Elements of Statistical Learning -- Data Mining, Inference, and Prediction}.
%   New York: Springer, 2009.
% \bibitem[4]{YourName17-XXX}
%   F.\ Lastname1, F.\ Lastname2, and F.\ Lastname3,
%   ``Title of your INTERSPEECH 2018 publication,''
%   in \textit{Interspeech 2018 -- 19\textsuperscript{th} Annual Conference of the International Speech Communication Association, September 2-6, Hyderabad, India Proceedings, Proceedings}, 2018, pp.~100--104.
% \end{thebibliography}

\end{document}